# In Brain Multi-Photon Imaging of Vaterite Drug Delivery Cargoes loaded with Carbon Dots

Hani Barhum[1,2,3,*,=], Cormac McDonnell[1,3,=], Oleksii O. Peltek[4], Rudhvi Jain[5], Galit Elad-Sfadia[5], Muhammad Athamna[2,5] Pablo Blinder[5,6], and Pavel Ginzburg[1,3]

[1]Department of Electrical Engineering, Tel Aviv University, Ramat Aviv, Tel Aviv 69978, Israel

[2]Triangle Regional Research and Development Center, Kfar Qara'3007500, Israel;

[3]Light-Matter Interaction Centre, Tel Aviv University, Tel Aviv, 69978, Israel

[4]School of Physics and Engineering, ITMO University, St. Petersburg 191002, Russian Federation

[5]Neurobiology, Biochemistry and Biophysics School, Wise Life Science Faculty, Tel Aviv University, Tel Aviv, 69978, Israel

[6]Sagol School of Neuroscience, Tel Aviv University, Tel Aviv, 69978, Israel

*corresponding author
=equal contribution

Biocompatible fluorescent agents, such as phenylenediamine carbon dots (CDs), are key contributors to the theragnostic paradigm, enabling real-time in vivo imaging of drug delivery cargoes. This study explores the optical properties of these CDs, demonstrating their potential for two-photon fluorescence imaging in brain vessels. Using an open aperture z-scan technique, we measured the wavelength-dependent nonlinear absorption cross-section of the CDs, achieving a peak value near 50 GM. This suggests the potential use of phenylenediamine CDs for efficient multiphoton excitation in the 775 - 895 nm spectral range. Mesoporous vaterite nanoparticles were loaded with fluorescent CDs to examine the possibility of a simultaneous imaging and drug delivery platform. Efficient one and two-photon imaging of the CD-vaterite composites, uptaken by macrophage and genetically engineered C6-Glioma cells, was demonstrated. For an in vivo scenario, vaterite nanoparticles loaded with CDs were directly injected into the brain of a living mouse, and their flow was monitored in real-time within the blood vessels. The facile synthesis of phenylenediamine carbon dots, their significant nonlinear responses, and biological compatibility show a viable route for implementing drug tracking and sensing platforms in living systems.

## 1. Introduction

Biocompatible fluorescent agents, such as carbon dots (CDs), have become a cornerstone in theranostics, enabling real-time in vivo imaging of drug delivery cargoes [1,2]. CDs have attracted considerable attention across various disciplines, including bioimaging [3], biosensing [4], and biotherapy [5,6] due to their numerous advantages. CDs can be synthesized through facile production routes in wet chemistry and physical processes [7], including laser ablation [8], microwave-assisted methods [9], and plasma processing [10]. Their tunable fluorescence [11], high quantum yield [12] and various other physicochemical properties of CDs are related to the carbon source [13] and synthesis protocols [14], including parameters such as reaction temperature [15], time [16], pH [17], and post-synthetic treatments [18].

The potential biocompatibility of CDs compared to common toxic inorganic crystalline quantum dots [19] and other organic dyes [20,21], places them at the forefront of bio-imaging and bio-sensing agents [22]. Moreover, CDs can be modified to act as targeted binders [23,24]. This means they can be engineered to recognize and bind to specific sites on cells or tissues, enhancing their utility in targeted imaging [25]. This targeted binding not only aids in delivering the CDs to the desired location but also makes the site visible or trackable, which is particularly useful in the real-time monitoring of biological processes [26]. Several types of CDs exhibit outstanding nonlinear responses in multiphoton fluorescence (MPF) [27]. As such, many studies have been undertaken with implementing CDs in deep tissue imaging and therapy [28,29]. For instance, nitrogen-doped CDs have been developed for high-resolution imaging and ultrasensitive sensing of metal ions [30], while two-photon radiometric carbon dot-based nanoprobes have been used for real-time monitoring of intracellular pH [31]. Polyphenolic carbon quantum dots have also been employed for membrane-targeting and drug delivery in tumor therapy [32]. Furthermore, the nonlinear response of CDs can be exploited for photodynamic therapy [33], where the energy from the absorbed photons is used to generate reactive oxygen species to kill cancer cells. Among the applicable CDs utilized in diverse applications are multicolored Phenylenediamine CDs [30,34]. These colorful bright CDs showed imaging, sensing, and therapy abilities making them particularly attractive for further theranostic development.

Theranostics combines diagnostics and therapy in a single platform, necessitating drug-delivery nanocapsules to possess multiple functions, including trackability with high spatial resolution [35]. Concerning CDs, it can be noted that their application in theranostics is mostly realized in combination with other materials [36]. In this context, mesoporous calcium carbonate ($CaCO_3$) in vaterite form has emerged as an exceptionally promising load-carrying particle. This non-organic nanoparticle offers several advantages, including a high load capacity, large surface area, and facile chemistry for ligand binding, allowing it to accommodate different types of cargo [37–39]. Furthermore, its biodegradability makes it a safe option for in vivo applications [40,41]. The potential integration of CDs with vaterite nanoparticles combines the imaging capabilities of CDs with the drug delivery potential of vaterite, creating a powerful platform for theranostics. Finally, despite considerable research efforts dedicated to developing nanoparticle-based drug delivery systems for cancer treatment, the efficiency of nanoparticle accumulation at the tumor site often falls short, rarely exceeding a few percent [42]. This limitation underscores the importance of in vivo studies to optimize nanoparticle properties for a more efficient drug delivery process [43].

This manuscript explores multiphoton imaging of vaterite nanoparticles loaded with phenylenediamine CDs, as schematically shown in Figure 1. The imaging ability of the CD-vaterite composite is demonstrated through single and multiphoton fluorescence microscopy of cells and in vivo brain vessels. The CDs are characterized by several methods, including Fourier Transform Infrared (FTIR) Spectroscopy, X-ray

Photoelectron spectroscopy, and Transmission Electron Microscopy (TEM). The two-photon absorption cross-section of phenylenediamine CDs in biological conditions was characterized and compared to Fluorescein. The CDs were then loaded onto vaterite nanoparticles and used for one and two-photon imaging of Glioma cells. Finally, the CD-vaterite composites were imaged in vivo in brain vessels, providing a powerful tool for studying drug-cell interactions. As an outlook, the demonstrated approach contributes to developing multi-functional optical tools for drug-cell interaction studies.

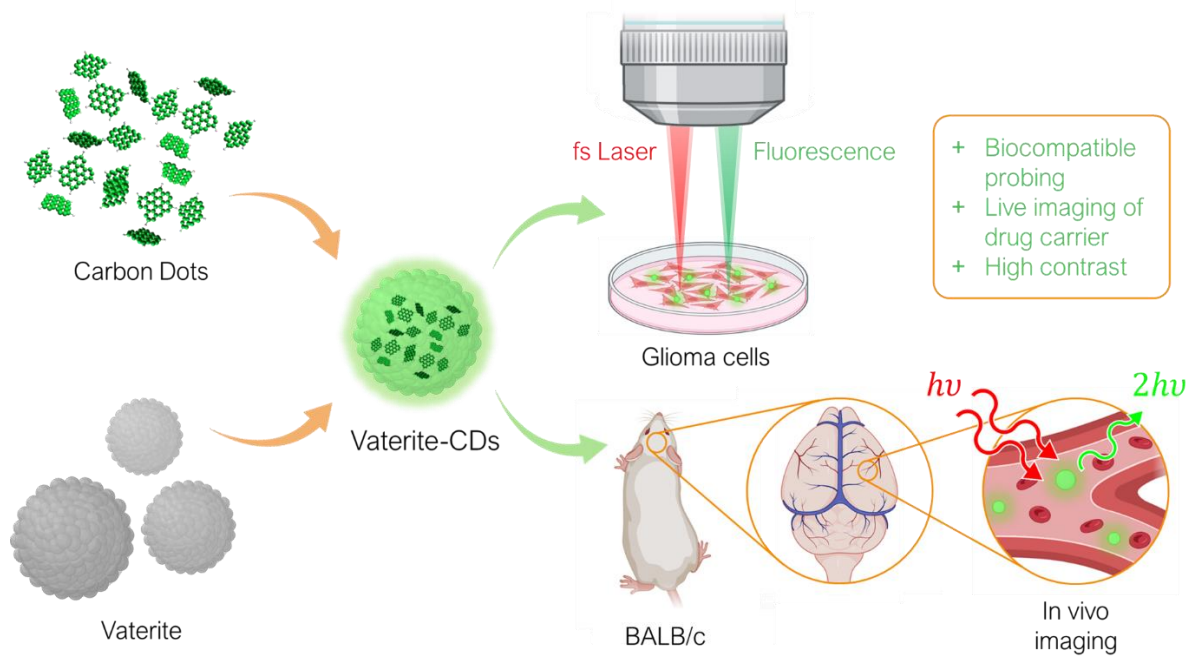

Figure 1. The Concept. (Left) CD synthesis followed by mesoporous vaterite encapsulation. (Middle) Vaterite particles loaded with CDs. Illustration of Glioma cells, incubated with fluorescent particles excited with a NIR femtosecond laser. (Bottom-right) Illustration of in vivo imaging of CD-vaterite composites in brain blood vessels.

## 2. Methods

2.1 Materials

All the chemicals were purchased from Sigma Aldrich and were used without further purification. Meta-phenylenediamine (mPD) >99%, ethylene glycol 99.8%, hydrochloric acid 32% (HCl), acetonitrile (AcN) (AR), ethanol (AR), $CaCl_2$ 99.9%, $Na_2CO_3$ 99.9%, $NaHCO_3$ 99.9%, poly-styrene sulfonate *70,000 Mw* (99.9%), fluorescein 98%, fluorescein isothiocyanate (FITC)>90%, MilliQ water, poly allylamine (PAH).

2.2 CD synthesis

Fluorescent CDs nanoparticles were synthesized using meta phenylenediamine as the carbon source. Detailed information about CD synthesis appears in our previous publication[30]. Briefly, the solvothermal reaction was performed in an ethylene glycol medium and acidified by hydrochloric acid. 95 mL of ethylene glycol and 5 mL concentrated HCl solution in three-necked flasks were heated to $453^0 K$. Then 1 gram of meta phenylenediamine was added and refluxed for 3 hours. The dark solution is then cooled to room temperature. In the following step 1:5 ratio of carbonization extract and acetonitrile was mixed and centrifugation at 5 krpm for 30 min. The supernatant was transferred to a 100 mL beaker and then removed by heating to $353^0 K$. The remaining concentrated CDs dark solution was dried on a thin glass and then washed into a 100 mL beaker with ethanol. The solution was dried overnight in a vacuum chamber. The last steps of purifying were repeated to achieve pure CDs for analysis.

2.3 Vaterite Synthesis

The vaterite synthesis was performed according to a previously developed protocol. Briefly, the general reaction conditions were 85:15 ethylene glycol to water ratio in a 40 mL Erlenmeyer, and the rotation speed of the magnet bar was set to 400 rpm at room temperature. CaCl2 0.025 M and Na2CO3 0.005 M concentrations were mixed for 120 min to achieve spherical particles. Elliptical particles were synthesized from $CaCl_2$ 0.005 M and $NaHCO_3$ 0.025 M, mixing for 120 min. Toroidal particles were synthesized with the same spherical particle methodology synthesis conditions with the addition of 24 mg of polystyrene sulfonate before the addition of $CaCl_2$. The particles are collected using centrifugation at 5 krpm for 30 min. The particles were washed with EtOH 5 ml three times before use.

2.4 Structural analysis

The size of the CD nanoparticles was examined using High-Resolution Transmission Electron Microscopy (HR-TEM) at an acceleration voltage of 200 kV. A Liquid Chromatography Mass Spectrometer (LCMS) was used to analyze the CD fragment mass. Fourier Transform Infrared spectroscopy (FTIR) was employed to extract the characteristic chemical groups of the CDs. Scanning Electron Microscopy (SEM) was employed to visualize vaterite particles.

2.5 CD-Vaterite Composites

Purified and dried vaterite particles 20 mg were vortexed and sonicated for 5 mins every half hour with CD solutions of 2 mL at 2-4 mg/mL solution in EtOH. After two hours of adsorption, the solution was centrifuged at 15 krpm for 10 mins. The remaining CD solution was removed, and the particles were washed twice by centrifugation in EtOH or DIW, depending on the measurement requirements.

2.6 Linear and Nonlinear Optical Characterization

Linear absorbance, emission, and photoluminescence excitation spectroscopy (PLE) were performed with a plate reader Synergy H1. Confocal microscopy images were obtained using a Leica SP8 system with 488 nm diode laser. Nonlinear fluorescence was investigated using an optical parametric oscillator (OPO) laser (Coherent Chameleon), with excitation wavelengths from 780 *to* 890 nm. The system generates 200 fs pulses at a repetition rate of 80 MHz. The laser output was focused into a cuvette holding the CDs with a microscope objective, and the reflected excitation was collected using the same objective. The reflected light was analyzed using a spectrometer after filtering the pump beam.

2.7 Quantum Yield, Two-Photon Fluorescence Cross Section, and Two-Photon Absorbance

The Quantum Yield (QY) of CDs was measured in comparison to fluorescein and rhodamine 6G. The standard slope method was used. The obtained QY values are in the range ≈ 40 %, corresponding to the previously reported data[44]. To quantify the absorption cross-section and the multi-photon cross-section fluorescein dye molecules were used as a reference. The two-photon absorption coefficient was measured using the open aperture Z-scan technique. The cross-section equation is given by:

$$\delta = \delta_r \times \frac{\Phi_r}{\Phi} \times \frac{F}{F_r}\frac{C_r}{C} \qquad (1)$$

where $\delta_r$ is the known fluorescein cross-section, $\Phi$ is CD QY, $\Phi_r$ is the relative QY used for fluorescein, $F$ is the integrated fluorescence, $n$ is the refractive index of the medium, and $C$ is the molarity. The nonlinear absorption coefficient in the Z-scan technique is given by:

$$\Delta T(z) = \sum_{m=0}^{\infty} \frac{[-q_0(z,0)]^m}{(m+1)^{3/2}} \quad q_0 = \frac{\beta I_0 L_{eff} z_0^2}{z^2 + z_0^2} \qquad (2)$$

where the nonlinear absorption $\beta = \frac{2\sqrt{2}}{I_0 L_{eff}}\Delta T$. T is the transmission from the sample, $L_{eff}$ is the sample thickness, $I_0$ is the initial transmitted power far from the focal point. The data was fitted numerically [45].

2.8 Cell Culture and Imaging

The Glioma cells (GL261) were modified to constitutively express tdTomato. Cells were cultured in DMEM with 10% FBS and maintained at 37°C in a humidified 5% $CO_2$ incubator. Cells were plated in a 35 mm dish with a density of 5 X 105 cells/dish for 2 days until used in the imaging experiments. The cells were then loaded with the CD embedded in vaterite through incubation of 0.1 mg/mL for 2 hrs. Followed by washing and replacement with a cell growth medium. Imaging was conducted in an upright multiphoton imaging microscope (Sutter MOM, USA) equipped with a tunable laser source emitting 140 fs pulses at a repetition rate of 80 MHz (Chameleon Discovery NX laser, Coherent Inc). A 750 nm long-pass dichroic was used to direct fluorescent emitted photons into a collection arm consisting of a 535 ± 20 nm bandpass and a photomultiplier tube (Hamamatsu). The microscope was controlled through Scan Image (Vidrio Technologies Inc, USA).

2.9 Animals and in vivo imaging

In vivo studies utilized healthy male C57BL/6J wild-type (WT) mice, aged 8 weeks and weighing 20-22 g, obtained from The Jackson Laboratory. The mice were housed in plastic cages within a specific pathogen-free environment maintained at 22 °C and a 40-60% humidity range. The mice were housed in a 12 h light/dark cycle and had ad libitum access to pellet food and water. The authorities of Tel Aviv University approved the experimental procedures for animal use and welfare and fully complied with IACUC guidelines.

The mouse was anesthetized using 5% inhalant isoflurane for induction and maintained at 1-2% in a 30/70 oxygen/N2O mixture throughout the surgical process. Following anesthesia, the mouse was placed onto a stereotaxic frame and the core body temperature was monitored and maintained at 37 °C using an electric heating pad. Ear bars and an incisor bar set suitable for a fat skull (3-4 mm) were used to secure the mouse. Before proceeding, eye lubricant was applied to prevent corneal desiccation. The mouse received an injection of dexamethasone (intramuscular, 0.2 mg/kg) to prevent swelling of the brain and/or inflammatory response and carprofen (subcutaneously, 5 mg/kg) for analgesia. Following these preparations, the skin surrounding the skull was cleaned using an isopropyl alcohol swab, and the scalp was cautiously removed and cleaned of any surrounding tissue or hair to expose the skull. The bone was then slowly drilled with a high-speed manual drill (Osada, drill bit #5) equipped with a round engraver drill bit was used for the subsequent craniotomy. The drilling process involved frequent pauses to allow for cooling of the skull bone using artificial cerebrospinal fluid (ACSF). With the aid of tweezers, the skull piece was carefully detached from the skull, exposing the underlying brain tissue[46,47]. The CD-vaterite composites were injected into the tissue using a manual microinjection system. The CD-vaterite composites were injected through a glass capillary tube approximately 100 µm deep under the Pia mater (1 µL, 1 mg/mL). The injections were performed at three different sites. After the injection, the cranial window was covered with #0 thickness cover glass for optical access. Then, a custom-made metal head frame was attached using cyanoacrylate glue (Loctite 401) and dental cement (high-Q-bond, BJM labs) to enable head fixation during imaging. To visualize the blood flow, mice were injected with 5% Texas Red dissolved in a CSF (25 µL, retro-orbitally). Imaging was conducted using the microscope mentioned above.

## 3. Results

Phenylenediamine, characterized by its two amine groups, plays a crucial role during the carbonization process, acting as a hot spot. The growth of the CDs is regulated on two levels. Firstly, the solvent ethylene glycol constrains the diffusion of reactive species. Secondly, adding an acid or proton source (in this case, 1%) shields the hot spot, resulting in kinetic control over the final product. This dual-level regulation directly influences the quantity of the product, its band gap, and the fluorescence quantum efficiency[30]. In our study, the CDs were grown to a diameter of approximately 3-5 nm. Figure 2 (a) shows a representative Transmission Electron Microscopy (TEM) scan of particles dispersed on a copper grid. The statistical analysis of nanoparticle sizes is depicted in Figure 2 (a) inset. The relatively small size of the CDs is advantageous for optimal loading into the mesoporous vaterite volume. Figure 2 (b) displays the Fourier Transform Infrared (FTIR) absorbance, identifying functional chemical groups. These groups are associated with fluorescence, chemical binding, and crosslinking, crucial for achieving specific site imaging. The FTIR spectrum reveals several primary peaks. The peak at 613 cm$^{-1}$ is associated with -CH2 rocking. Peaks around 900 cm$^{-1}$ correspond to the out-of-plane stretching of the aromatic -CH bonds. The peak at 2950 cm$^{-1}$ corresponds to a -CH stretching vibration, while the one around 1600-1700 cm$^{-1}$ corresponds to C=O and C=C bonds, respectively. The peak at 1100-1200 cm$^{-1}$ represents the N-C O-C stretching in the molecules. The pronounced peak at 3650 cm$^{-1}$ corresponds to -OH hydroxyl group stretching, and the band around 3200 cm$^{-1}$ is related to the -NH group of the primary and secondary amines.

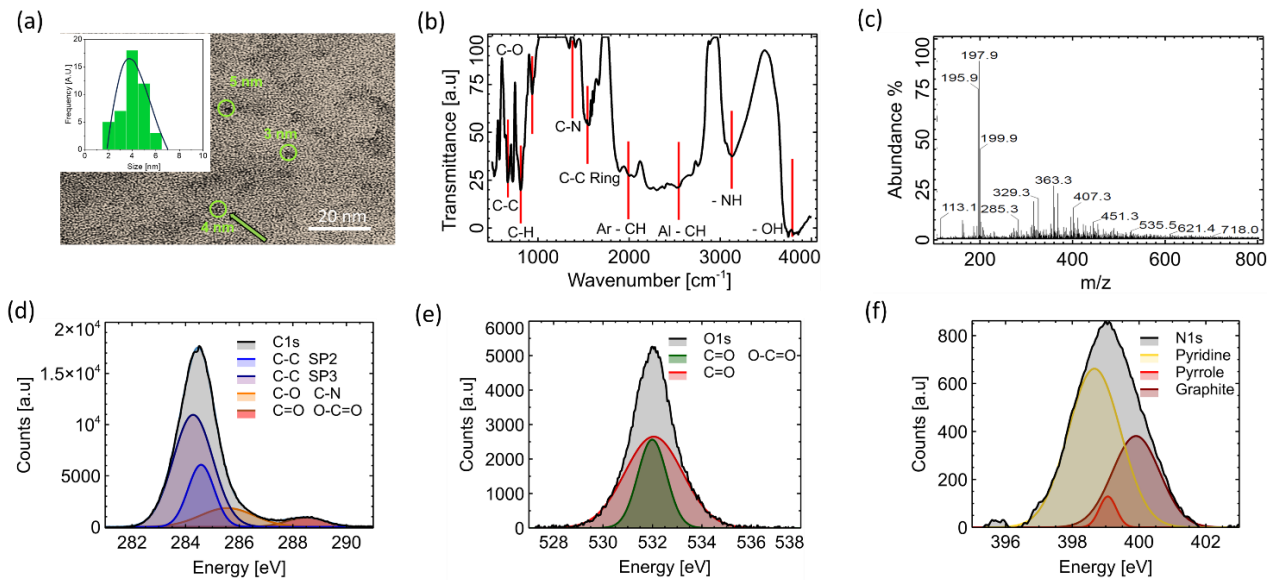

Figure 2: Characterization of the synthesized CDs. (a) Transmission Electron Microscopy (TEM) image of the CDs, with the inset showing the size distribution of the CDs. (b) Fourier Transform Infrared (FTIR) spectroscopy of the CDs, with red lines indicating the chemical fingerprints of the CDs. The FTIR spectrum provides insights into the various functional groups present on the CDs. (c) Liquid Chromatography-Mass Spectrometry (LC-MS) of the CDs further confirms the CDs' chemical composition. (d) X-ray Photoelectron Spectroscopy (XPS) of C 1s, (e) O 1s, and (f) N 1s, revealing the presence of various functional groups on the CDs' surface.

The Liquid Chromatography-Mass Spectrometry (LCMS) spectrum shown in Figure 2 (c) presents the mass distribution of the CD fragments. The primary fractions are distributed around 400 m/z, mass-to-charge, meaning that the primary and stable fragments are within this mass distribution. reinforcing the evidence that the CDs are small. The XPS analysis, as shown in Figure 2 d-f, provides a detailed view of the surface chemistry of the CDs. The C 1s spectrum reveals several sub-peaks, indicative of various carbon bonds such as C-N, C-O, and C=N, demonstrating the versatility of CDs' surface chemistry. The presence of carbonyl or carboxylic acid groups (C=O), as indicated by the peak at ~287.8 eV, contributes to the CDs' dispersibility in aqueous solutions. The peak at approximately 288.5 eV points to carboxylic acid (COOH) groups, demonstrating the CDs' exceptional interfacial interaction capacities. The O1s spectrum further confirms the presence of oxygen-rich groups, contributing to the CDs' hydrophilicity and water stability. The N1s spectrum reveals the presence of pyridinic N, pyrrolic N, and graphitic N, which are integral to the fluorescence and electronic properties of the CDs. Overall, the XPS analysis underscores the complex interplay of functional groups that contribute to the notable properties of CDs, such as enhanced fluorescence and high water stability.

Figure 3 (a) illustrates the CDs' UV-VIS absorbance. The CDs exhibit two main absorbance bands—the band at 450 nm results from surface states and surface oxidation. The UV band around 270 nm arises due to the π→π* and n→π* transitions associated with conjugated C=C and C=O bonds. An emission band is observed at 525 nm. Prior research suggests this emission is likely due to surface defects on the CDs.

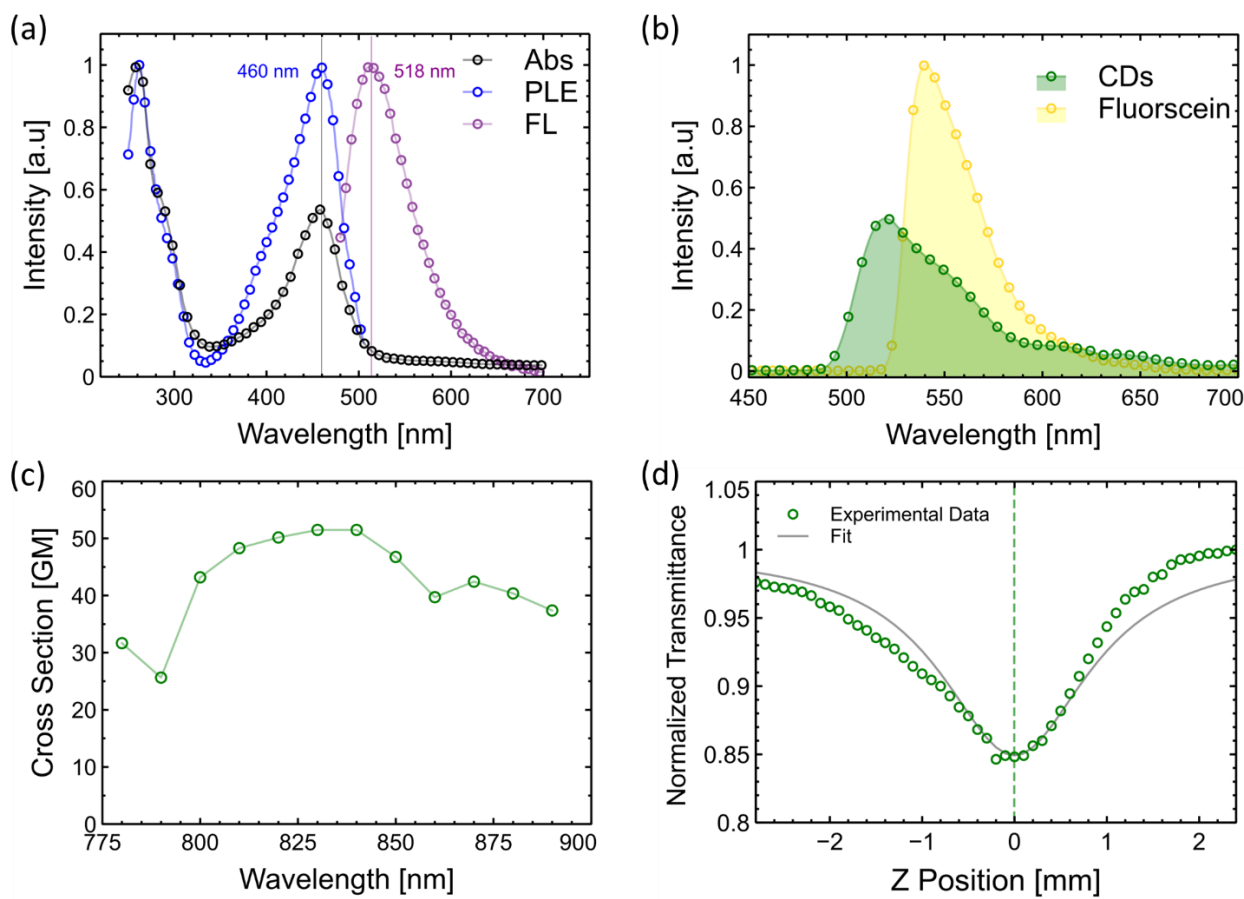

Figure 3. Linear and nonlinear properties of CDs. (a) Single-photon absorbance (represented by black circles), single-photon luminescence excitation (blue) monitored at 530 nm, and single-photon fluorescence (green) under 450 nm excitation spectra. (b) Two-photon fluorescence spectra of CDs (represented by green circles) and Fluorescein (yellow) under 800 nm femtosecond pump irradiation. The graphs are to scale. (c) Two-photon absorption cross-section of CDs (calibrated to Fluorescein) versus pump wavelength. (d) Open aperture Z-scan of CDs (0.03 mg/mL, in an ethanol solution) in a 2 mm quartz cuvette.

The two-photon fluorescence spectra of the CDs and Fluorescein are shown in Figure 3 (b), following excitation with an 800 nm femtosecond pump. The emission band of the CDs is broader than Fluorescein and extends from green 490 nm – 600 nm with a lower intensity up to 650 nm. The normalized intensity across the CDs spectrum, which has a Quantum Yield (QY) of 35%, is approximately 0.6 when compared to Fluorescein, which has a QY of 91%. The two-photon cross-section was calculated relative to Fluorescein, which served as a reference standard. This calculation was performed for the 780 - 900 nm wavelength range using Equation (1), and the results are displayed in Figure 3 (c). The cross-section in Goeppert-Mayer (GM) units is approximately 30 at 780 nm excitation and above 50 GM at 830 nm. Beyond this point, the value decreases to 50 GM and oscillates between 50-30 GM with increasing wavelength. For comparison, the Fluorescein cross-section is 40 GM around wavelengths of 800 nm. The nonlinear two-photon absorbance of the CDs was measured using the open aperture Z-scan method[48]. An average laser power of 120 mW illuminated the sample, with a CD concentration of 0.01mg/mL in Ethanol. The solution was placed in a 2 mm thick quartz cuvette. Using Equation (2), a fitting with first order (m=1) in the z-scan transmittance was sufficient. The CD's two-photon absorption coefficient is estimated to be 2 cm·W$^{-1}$ from the fit. In comparison, several other materials, including organic molecules like coumarin [49], bis(styryl)benzene derivatives [50], and quantum dots [51], exhibit extremely high absorption coefficients ranging from 100 to 10,000 GM.

In the next phase, the CDs were encapsulated within vaterite nanoparticles. This process was aimed as a first step towards creating CD-vaterite composites that could perform dual functions of both drug delivery and imaging. We fabricated vaterite cargo particles of varying geometries and immersed them in CD solutions of different concentrations. The SEM images in Figure 4 (a,b,c) show the precise geometry and porosity of the vaterite particles before encapsulation. The CD-vaterite composites were then imaged using a confocal microscope, measuring approximately 500 nm for individual particles and in different shapes. Those were chosen to have a similar volume for all particles. More statistical info from SEM data and confocal images are presented in the *Supporting file*. The statistical images are presented in Figure 4 (d,e,f). Although the resolution was not high enough to determine the exact concentration of the CDs within the volume, a general trend could be identified. Moreover, the toroid particles exhibited greater brightness, suggesting a higher presence of fluorescent agents on the surface since there is no significant loading of CDs relative to other shapes. The fluorescent image of the toroid also mirrors its donut-shaped topology. Data from ten particles were collected, and the average distributions are displayed in the panels, with error bars represented by vertical lines. The fraction of CDs adsorbed on vaterite was quantified using the Langmuir parameters and loading properties. Figure 4 (g) shows the Langmuir adsorption curves of CD molecules on the vaterite surface, plotted against the CD concentration in the solution during fabrication. The adsorption reached saturation between 0.7 – 0.9 mg·g$^{-1}$, with elliptical particles achieving the highest CD load capacity—linearization as described in the supporting file, and equation S1 of the Langmuir equation enabled the extraction of the isotherm parameters seen in (Table 1). In general, despite the differences in shape properties, the number of CDs loaded on a particle is only weakly dependent on its shape, as shown in Figure 4 (h). The Langmuir adsorption parameters further elucidate this observation. The Langmuir constant *K*, which reflects the affinity between the CDs and the particles, remains relatively consistent across different particle geometries. This suggests that the inherent binding energy or adhesion energy between the CDs and the particles is similar, irrespective of the particle's shape. However, the maximum adsorption capacity $q_m$ does exhibit variations, with elliptical particles there is a slightly higher capacity. This could be attributed to the unique surface properties or internal structure of the elliptical particles, which might offer more favorable sites for CD adsorption.

Table 1. Langmuir Isotherm parameters for different shapes of Vaterite.

| Geometry | $q_m \left(\frac{mg}{g}\right)$ | $K \left(\frac{L}{mg}\right)$ |
|---|---|---|
| Spherical | 0.183 | 61.98 |
| Elliptical | 0.209 | 62.99 |
| Toroidal | 0.176 | 63.74 |

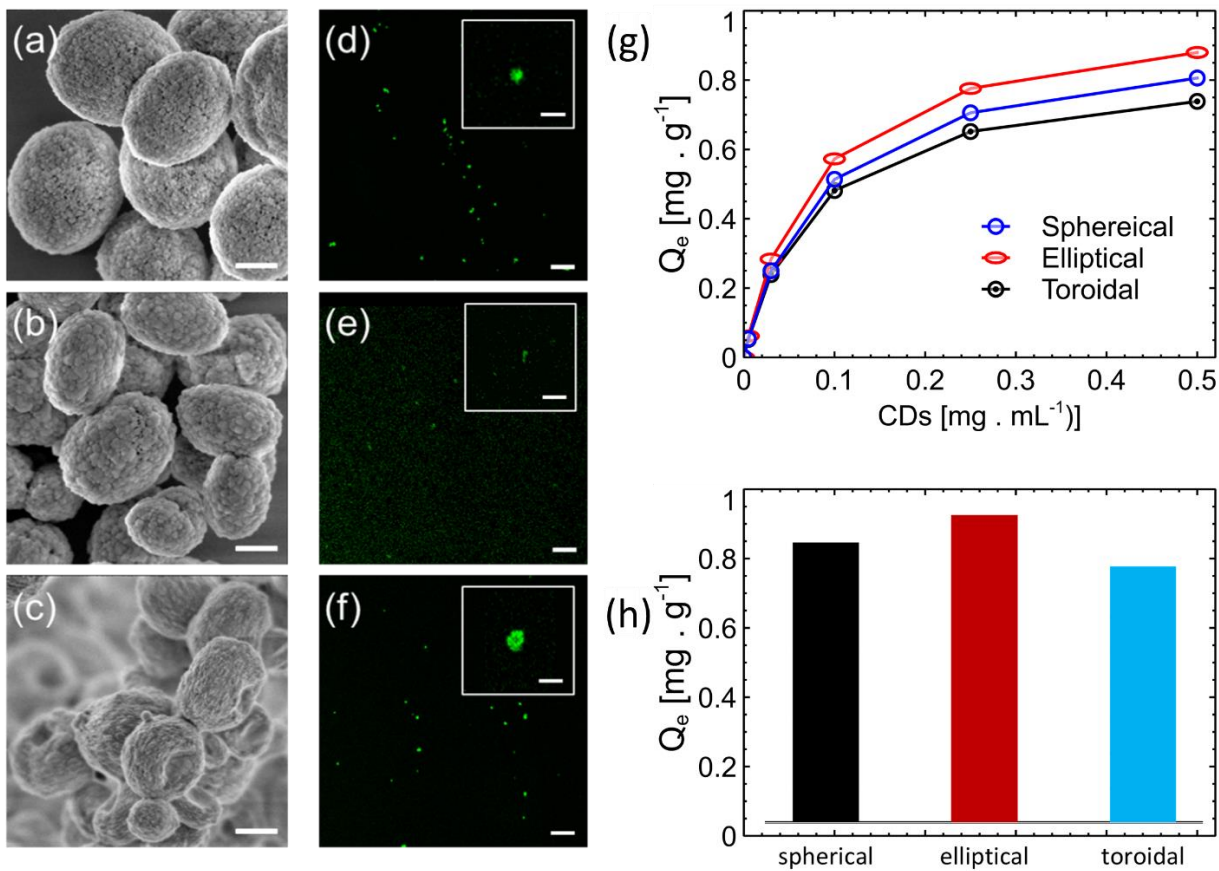

Figure 4. (a,b,c) Scanning electron microscope images of vaterite particles in three different geometries: (a) spheres, (b) ellipsoids, and (c) toroids. The scale bar is 400 nm. (d,e,f) Confocal one-photon fluorescent microscope images of vaterite particles, loaded with CDs, in the three different geometries: (d) spheres, (e) ellipsoids, and (f) toroids. The insets provide a closer view of a single particle with a scale bar of 0.5 µm. (g) Langmuir adsorption curve depicting the adsorption of CDs on a vaterite surface for different particle geometries. (h) Maximum adsorption of the CDs on vaterite for different geometries, as derived from the Langmuir equation.

Several in vitro experiments were conducted to show the CDs' imaging potential. Macrophages, which are white blood cells integral to the innate immune system, are known for their ability to efficiently uptake foreign objects in their environment. This cell type was chosen to test the CDs' one- and two-photon imaging ability. The macrophages were stained by immersing them in a CD solution for 1 minute, followed by a washing process. The confocal image, as shown in Figure 5 (a), demonstrates the successful staining of the entire cell volume.

In the subsequent stage, CD-stained macrophages were mixed with spherical vaterite. Another set of vaterite nanoparticles was loaded with Rhodamine isothiocyanate, a red-emitting compound. Figure 5 (b) presents the overlaid confocal and bright image of CD-stained macrophages and Rhodamine-stained vaterite nanoparticles. The CD-vaterite composites, which were deposited on the substrate and subsequently taken up by the macrophage, can be clearly identified.

Next, the non-specific binding of CDs to C6-Glioma cells was demonstrated. A 2 mL cell culture with 100 µL of a 0.1 mg/mL CD solution was prepared. After 10 minutes of immersion, the sample was examined under the confocal microscope. Figure 5 (c) displays the staining ability of the CDs. To demonstrate the potential tracking ability of future treatments of glioma cells with vaterite nanoparticles, we used genetically engineered C6-Glioma cells that constitutively express td-tomato in the next set of experiments. The cells were incubated with CD-vaterite composites. After 30 minutes, the medium was washed to remove unbound vaterite particles. Due to their accelerated metabolic activity, cancer cells uptake vaterite particles quite efficiently, as shown in previous studies[37]. The C6-Glioma cells with uptaken CD-vaterite composites were imaged using two-photon microscopy. The td-tomato (red) was observed as a dynamic medium inside the cells. Figure 6 (a) is the two-photon fluorescence (TPF) image of the cell without CD-vaterite composites added. The fluorescent signal comes from tdTomato, which emits at 581 nm. Figure 5 (d) shows the cells after being uptaken with CD-vaterite composites. The vaterite particles emit green light at 520 nm. Panel (c) shows a zoomed image of the excitation region, where the vaterite particles inside the cell can be clearly seen.

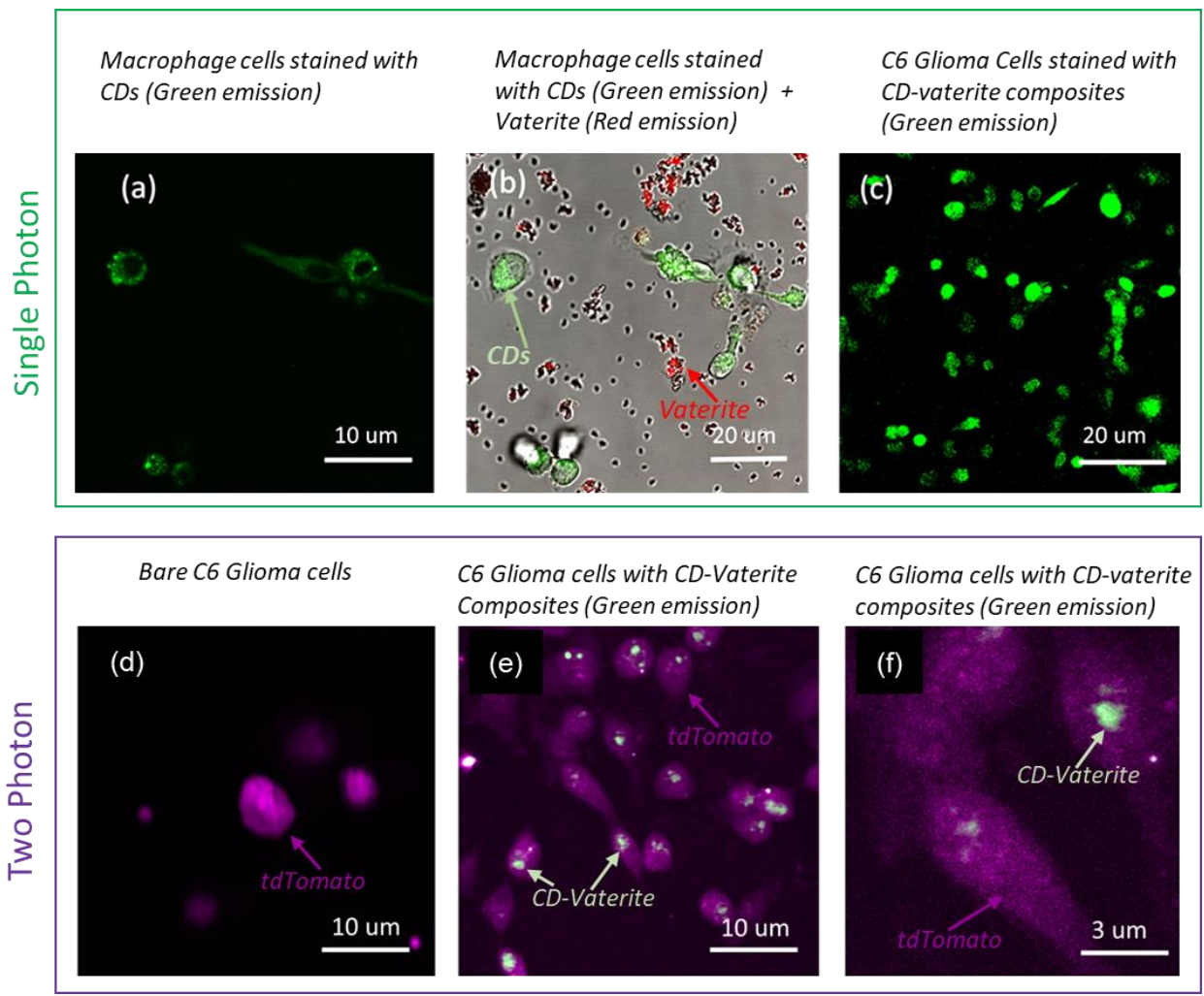

Figure 5. (a) Confocal microscope image of macrophage cells under 488 nm CW laser excitation with a 40x objective lens. (b) Combined confocal and bright field microscope image of Macrophage cells incubated with CDs under 488 nm CW laser excitation and Vaterite particles with Rhodamine isothiocyanate under

552 nm CW laser excitation with a 40x objective lens. (c) One-photon fluorescence confocal image of C6-Glioma cells embedded with a vaterite-CD cargoes. (d,e,f) TPF images of C6-Glioma cells under 800 nm excitation with a 10x immersion objective. (d) C6-Glioma cells without any vaterite added, the source of the 2PF is the tdTomato (e) C6-Glioma cells 2PF in expressed tdTomato, the green bright objects are CD-vaterite composites. (f) Single glioma cell with CD-vaterite composites emitting TPF.

Next, two-photon microscopy of C6-Glioma cells with uptaken CD-vaterite composites was performed. The genetically engineered C6-Glioma cells, which constitutively express tdTomato, provide a dynamic medium for imaging. The red tdTomato fluorescence is observed distinctly in the cells. Figure 5 (d) displays the TPF image of the cell without CD-vaterite composites. The fluorescent signal comes solely from tdTomato, which has a peak emission at 581 nm. This image provides a baseline for comparison with the cells that have uptaken the CD-vaterite composites. Figure 5 (e-f) presents the cells after the uptake of CD-vaterite composites. The vaterite particles emit green light at 520 nm, contrasting the red tdTomato fluorescence. This contrast allows for clearly visualizing the CD-vaterite composites within the cells. A zoomed image of the excitation region is shown in Figure 5 (f), where the vaterite particles inside the cell can be clearly seen. This detailed view further underscores the successful uptake of vaterite-CD particles by the C6-Glioma cells and the effectiveness of two-photon microscopy in visualizing these particles.

In the next step, particle visualization was tested in an in vivo setting. The process involved introducing spherical sub-micron CD-vaterite composites with microinjections into brain tissue through a cranial window, as depicted in Figure 6 (a, b). The blood plasma was stained with Texas Red to facilitate the visualization of blood vessels. The local injection site was imaged using two-photon microscopy. Figure 6 (d) displays multiple imaged time steps over 5.6 s, showing the bloodstream in red and the surrounding area. The vaterite CD particles are distinguishable from the background due to their green fluorescence emission. Individual vaterite particles can be observed flowing relatively fast in the bloodstream in several images. For a more detailed observation, additional frames show the dynamic movement of these particles in the bloodstream and are provided in Figure S2. Notably, there is also a green fluorescence observed, which can be attributed to the release of material from the vaterite particles. This study marks the first observation of vaterite particles in a mouse brain, underscoring the novelty of our approach.

While the CD-vaterite composite might not singularly represent the pinnacle in fluorescence or drug-carrying capabilities, its unique combination offers a biocompatible multitasking platform. This synergy between the fluorescence of CDs and the drug-carrying potential of vaterite emphasizes the promise of such composites in theranostic applications.

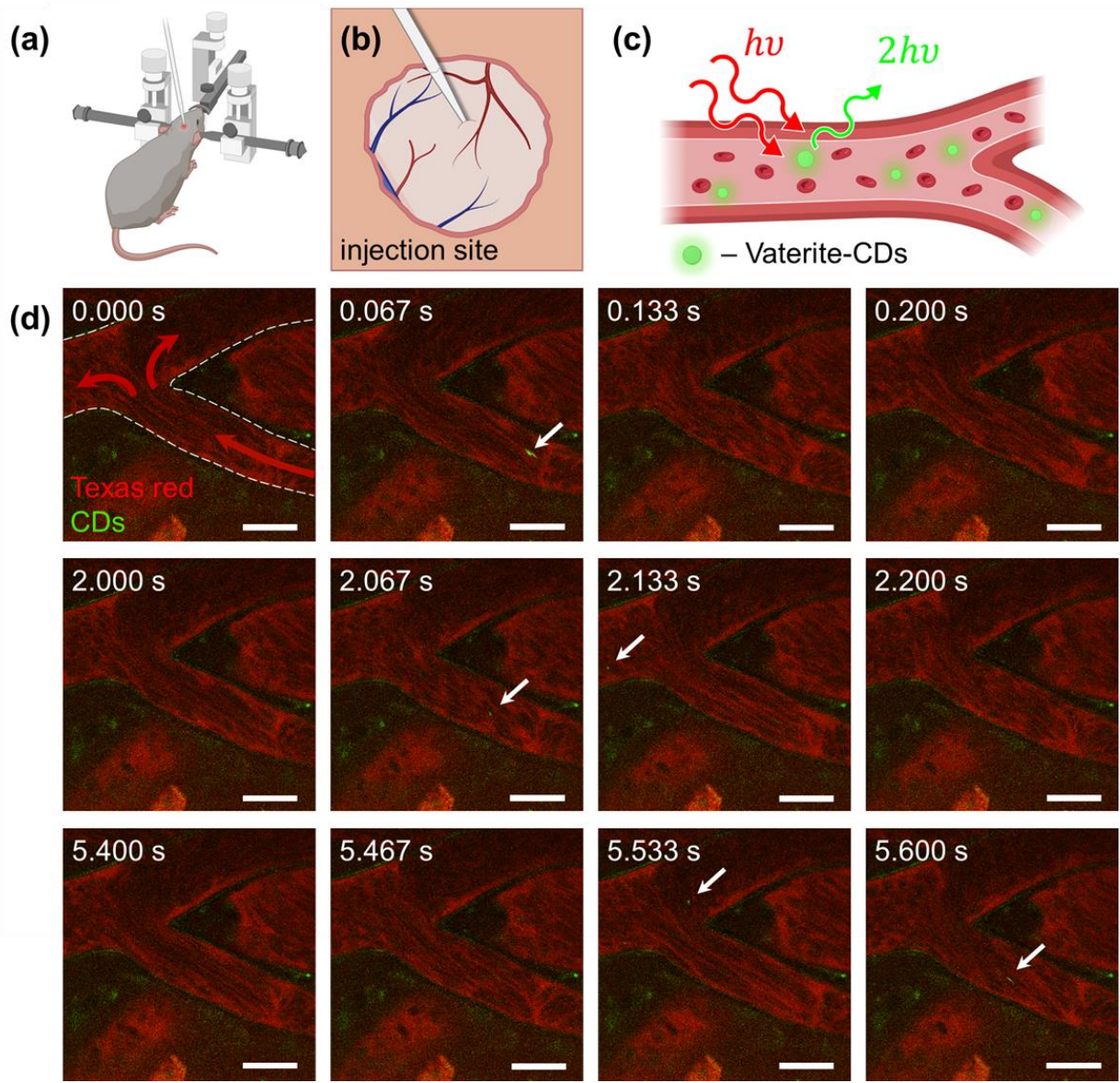

Figure 6. (a) Illustration of a mouse during the administration of CD-vaterite composites. (b) A schematic representation of the injection with CD-vaterite composites (c) A schematic of two-photon imaging used to visualize the CD-vaterite composites (d) Two-photon images capturing the presence of CD-vaterite composites in the bloodstream. The red color corresponds to Texas Red, employed to visualize the blood flow, while the green color represents the CD-vaterite composites. The white dashed line outlines the monitored blood vessel border. Red arrows indicate the direction of the blood flow, while white arrows highlight the CD-vaterite composites. The scale bar corresponds to 30 µm. Figures (a)-(c) were created with BioRender.com.

## 4. Discussion and Conclusion

In this study, we have demonstrated a facile synthesis of meta-phenylenediamine CDs, exhibiting robust optical fluorescence properties under both single-photon and two-photon absorption regimes. The detailed optical characterization of the CDs reveals a high two-photon cross-section across a wide range of pump wavelengths, comparable to the well-established staining compound Fluorescein. This underscores the potential of CDs as efficient and biocompatible nonlinear imaging agents. While other inorganic nonlinear materials, such as quantum dots, may have significantly higher fluorescent intensity and higher two-photon cross sections, CDs provide a biocompatible tool for imaging in cells and other organic materials. Biocompatibility plays a key role in continuous imaging of cells in the case of both in-vitro and in-vivo. This may enable long-term experiments with minimal side effects and experimental deviation in cell studies.

The CDs were loaded onto mesoporous vaterite nanoparticles in the next stage to demonstrate a complete biocompatible sample on pathways to a theranostic tool. The nonspecific binding of the vaterite surface to the CDs provides a straightforward CD loading route. Furthermore, this loading interaction could be further enhanced with specialized covalent linkers for specificity and is an outlook of this report.

The interaction of the CD-vaterite composites with macrophages and C6-Glioma cells was successfully investigated using one- and two-photon confocal microscopy. Following this, the CD-vaterite composites were directly visualized in-vivo in brain vessels for the first time to demonstrate, showing the ability to track single CD-vaterite composites flowing in the vessels. This proof of principle provides further guidelines to demonstrate a powerful tool for studying real-time drug-cell interactions. Also, it may be of specific benefit to study the rheological properties of brain circulation and the interaction of particles with the glycocalyx in capillaries. This work shows the powerful potential of CDs and vaterite as multifunctional tool that can be applied to multi-photon imaging and drug delivery.

**Acknowledgments**

This research was partially supported by the Naomi Foundation through the Tel Aviv University GRTF Program, ERC StG "In Motion" (802279), Tel Aviv University Breakthrough Innovative Research Grant, the Ministry of Science, Technology and Space of Israel (Grant No. 79518), and QuanTAU - Center for Quantum Science and Technology Equipment Grant and the Israeli Science Foundation (Grant no. 2342/21) and the European Research Council PoC (Grant no. 101066138).

**References**

1　Sharmiladevi P, Girigoswami K, Haribabu V, Girigoswami A. Nano-enabled theranostics for cancer. *Mater Adv* 2021; **2**: 2876–2891.

2　Ansari L, Hallaj S, Hallaj T, Amjadi M. Doped-carbon dots: Recent advances in their biosensing, bioimaging and therapy applications. Colloids Surf B Biointerfaces. 2021; **203**. doi:10.1016/j.colsurfb.2021.111743.

3　Zhao J, Li F, Zhang S, An Y, Sun S. Preparation of N-doped yellow carbon dots and N, P co-doped red carbon dots for bioimaging and photodynamic therapy of tumors. *New Journal of Chemistry* 2019; **43**: 6332–6342.

4   Liu L, Anwar S, Ding H, Xu M, Yin Q, Xiao Y *et al.* Electrochemical sensor based on F,N-doped carbon dots decorated laccase for detection of catechol. *Journal of Electroanalytical Chemistry* 2019; **840**: 84–92.

5   Geng B, Hu J, Li Y, Feng S, Pan D, Feng L *et al.* Near-infrared phosphorescent carbon dots for sonodynamic precision tumor therapy. *Nat Commun* 2022; **13**. doi:10.1038/s41467-022-33474-8.

6   Liu Y, Xu B, Lu M, Li S, Guo J, Chen F *et al.* Ultrasmall Fe-doped carbon dots nanozymes for photoenhanced antibacterial therapy and wound healing. *Bioact Mater* 2022; **12**. doi:10.1016/j.bioactmat.2021.10.023.

7   Yu C, Jiang X, Qin D, Mo G, Zheng X, Deng B. Facile Syntheses of S,N-Codoped Carbon Quantum Dots and Their Applications to a Novel Off-On Nanoprobe for Detection of 6-Thioguanine and Its Bioimaging. *ACS Sustain Chem Eng* 2019; **7**: 16112–16120.

8   Kaczmarek A, Hoffman J, Morgiel J, Mo´scicki TM, Stobí Nski L, Szymá Nski Z *et al.* materials Luminescent Carbon Dots Synthesized by the Laser Ablation of Graphite in Polyethylenimine and Ethylenediamine. *Materials* 2021; **14**. doi:10.3390/ma.

9   Kamali SR, Chen CN, Agrawal DC, Wei TH. Sulfur-doped carbon dots synthesis under microwave irradiation as turn-off fluorescent sensor for Cr(III). *J Anal Sci Technol* 2021; **12**. doi:10.1186/s40543-021-00298-y.

10  Ma X, Li S, Hessel V, Lin L, Meskers S, Gallucci F. Synthesis of luminescent carbon quantum dots by microplasma process. *Chemical Engineering and Processing - Process Intensification* 2019; **140**: 29–35.

11  Yang S, Sun J, Li X, Zhou W, Wang Z, He P *et al.* Large-scale fabrication of heavy doped carbon quantum dots with tunable-photoluminescence and sensitive fluorescence detection. *J Mater Chem A Mater* 2014; **2**: 8660–8667.

12  Tong L, Wang X, Chen Z, Liang Y, Yang Y, Gao W *et al.* One-Step Fabrication of Functional Carbon Dots with 90% Fluorescence Quantum Yield for Long-Term Lysosome Imaging. *Anal Chem* 2020; **92**. doi:10.1021/acs.analchem.9b05553.

13  Crista DMA, da Silva JCGE, da Silva LP. Evaluation of different bottom-up routes for the fabrication of carbon dots. *Nanomaterials* 2020; **10**: 1–15.

14  Lesani P, Lu Z, Singh G, Mursi M, Mirkhalaf M, New EJ *et al.* Influence of carbon dot synthetic parameters on photophysical and biological properties. *Nanoscale* 2021; **13**. doi:10.1039/d1nr01389k.

15  Xia C, Tao S, Zhu S, Song Y, Feng T, Zeng Q *et al.* Hydrothermal addition polymerization for ultrahigh-yield carbonized polymer dots with room temperature phosphorescence via nanocomposite. *Chemistry - A European Journal* 2018; **24**: 11303–11308.

16  Papaioannou N, Titirici MM, Sapelkin A. Investigating the Effect of Reaction Time on Carbon Dot Formation, Structure, and Optical Properties. *ACS Omega* 2019; **4**. doi:10.1021/acsomega.9b01798.

17   Tan C, Zhou C, Peng X, Zhi H, Wang D, Zhan Q *et al.* Sulfuric Acid Assisted Preparation of Red-Emitting Carbonized Polymer Dots and the Application of Bio-Imaging. *Nanoscale Res Lett* 2018; **13**: 272.

18   Craciun AM, Diac A, Focsan M, Socaci C, Magyari K, Maniu D *et al.* Surface passivation of carbon nanoparticles with: P -phenylenediamine towards photoluminescent carbon dots. *RSC Adv* 2016; **6**: 56944–56951.

19   Hu L, Zhong H, He Z. The cytotoxicities in prokaryote and eukaryote varied for CdSe and CdSe/ZnS quantum dots and differed from cadmium ions. *Ecotoxicol Environ Saf* 2019; **181**. doi:10.1016/j.ecoenv.2019.06.027.

20   Skanda S, Bharadwaj PSJ, Kar S, Sai Muthukumar V, Vijayakumar BS. Bioremoval capacity of recalcitrant azo dye Congo red by soil fungus Aspergillus arcoverdensis SSSIHL-01. Bioremediat J. 2021. doi:10.1080/10889868.2021.1984198.

21   Ismail M, Akhtar K, Khan MI, Kamal T, Khan MA, M. Asiri A *et al.* Pollution, Toxicity and Carcinogenicity of Organic Dyes and their Catalytic Bio-Remediation. *Curr Pharm Des* 2019; **25**. doi:10.2174/1381612825666191021142026.

22   Das R, Bandyopadhyay R, Pramanik P. Carbon quantum dots from natural resource: A review. *Mater Today Chem* 2018; **8**: 96–109.

23   Long C, Liu S, Li X, Zhu J, Zhang L, Qing T *et al.* In-situ covalent bonding of carbon dots on two-dimensional tungsten disulfide interfaces for effective monitoring and remediation of chlortetracycline residue. *Chemical Engineering Journal* 2022; **432**. doi:10.1016/j.cej.2021.134315.

24   Chen Y, Xiong G, Zhu L, Huang J, Chen X, Chen Y *et al.* Enhanced Fluorescence and Environmental Stability of Red-Emissive Carbon Dots via Chemical Bonding with Cellulose Films. *ACS Omega* 2022; **7**. doi:10.1021/acsomega.1c06426.

25   Chen B Bin, Liu ML, Li CM, Huang CZ. Fluorescent carbon dots functionalization. *Adv Colloid Interface Sci* 2019; **270**: 165–190.

26   Jiang K, Sun S, Zhang L, Lu Y, Wu A, Cai C *et al.* Red, green, and blue luminescence by carbon dots: Full-color emission tuning and multicolor cellular imaging. *Angewandte Chemie - International Edition* 2015; **54**: 5360–5363.

27   Cao L, Wang X, Meziani MJ, Lu F, Wang H, Luo PG *et al.* Carbon dots for multiphoton bioimaging. *J Am Chem Soc* 2007; **129**: 11318–11319.

28   Han Y, Liu H, Fan M, Gao S, Fan D, Wang Z *et al.* Near-infrared-II photothermal ultra-small carbon dots promoting anticancer efficiency by enhancing tumor penetration. *J Colloid Interface Sci* 2022; **616**. doi:10.1016/j.jcis.2022.02.083.

29   Lesani P, Singh G, Viray CM, Ramaswamy Y, Zhu DM, Kingshott P *et al.* Two-Photon Dual-Emissive Carbon Dot-Based Probe: Deep-Tissue Imaging and Ultrasensitive Sensing of Intracellular Ferric Ions. *ACS Appl Mater Interfaces* 2020; **12**. doi:10.1021/acsami.0c05217.

30    Barhum H, Alon T, Attrash M, Machnev A, Shishkin I, Ginzburg P. Multicolor Phenylenediamine Carbon Dots for Metal-Ion Detection with Picomolar Sensitivity. *ACS Appl Nano Mater* 2021; **4**: 9919–9931.

31    Lesani P, Singh G, Lu Z, Mirkhalaf M, New EJ, Zreiqat H. Two-photon ratiometric carbon dot-based probe for real-time intracellular pH monitoring in 3D environment. *Chemical Engineering Journal* 2022; **433**. doi:10.1016/j.cej.2021.133668.

32    Dutta SD, Hexiu J, Kim J, Sarkar S, Mondal J, An JM *et al.* Two-photon excitable membrane targeting polyphenolic carbon dots for long-term imaging and pH-responsive chemotherapeutic drug delivery for synergistic tumor therapy. *Biomater Sci* 2022; **10**. doi:10.1039/d1bm01832a.

33    Karagianni A, Tsierkezos NG, Prato M, Terrones M, Kordatos K V. Application of carbon-based quantum dots in photodynamic therapy. Carbon N Y. 2023; **203**. doi:10.1016/j.carbon.2022.11.026.

34    Sato R, Iso Y, Isobe T. Fluorescence Solvatochromism of Carbon Dot Dispersions Prepared from Phenylenediamine and Optimization of Red Emission. *Langmuir* 2019. doi:10.1021/acs.langmuir.9b02739.

35    Shailendrakumar AM, Tippavajhala VK. Gold and Carbon-Based Nano-theranostics: An Overview on the Developments and Applications for Cancer Phototherapy. Adv Pharm Bull. 2022; **12**. doi:10.34172/apb.2022.071.

36    Shen CL, Liu HR, Lou Q, Wang F, Liu KK, Dong L *et al.* Recent progress of carbon dots in targeted bioimaging and cancer therapy. Theranostics. 2022; **12**. doi:10.7150/thno.70721.

37    Bahrom H, Goncharenko AA, Fatkhutdinova LI, Peltek OO, Muslimov AR, Koval OYu *et al.* Controllable Synthesis of Calcium Carbonate with Different Geometry: Comprehensive Analysis of Particle Formation, Cellular Uptake, and Biocompatibility. *ACS Sustain Chem Eng* 2019; **7**: 19142–19156.

38    Volodkin D. CaCO3 templated micro-beads and -capsules for bioapplications. Adv Colloid Interface Sci. 2014; **207**. doi:10.1016/j.cis.2014.04.001.

39    Barhom H, Machnev AAA, Noskov RERE, Goncharenko A, Gurvitz EAEA, Timin ASAS *et al.* Biological Kerker Effect Boosts Light Collection Efficiency in Plants. *Nano Lett* 2019; **19**: 7062–7071.

40    Harpaz D, Barhom H, Veltman B, Ginzburg P, Eltzov E. Biocompatibility characterization of vaterite with a bacterial whole-cell biosensor. *Colloids Surf B Biointerfaces* 2023; **222**: 113104.

41    Parakhonskiy B, Svenskaya Y, Haase A, Lukyanets E, Antolini R. Anticancer drug delivery system based on vaterite particles. Journal of Biological Research (Italy). 2015.

42    Liu H, Pietersz G, Peter K, Wang X. Nanobiotechnology approaches for cardiovascular diseases: site-specific targeting of drugs and nanoparticles for atherothrombosis. J Nanobiotechnology. 2022; **20**. doi:10.1186/s12951-022-01279-y.

43	Biju V. Chemical modifications and bioconjugate reactions of nanomaterials for sensing, imaging, drug delivery and therapy. Chem Soc Rev. 2014; **43**. doi:10.1039/c3cs60273g.

44	Würth C, Grabolle M, Pauli J, Spieles M, Resch-Genger U. Relative and absolute determination of fluorescence quantum yields of transparent samples. *Nat Protoc* 2013; **8**: 1535–1550.

45	Mixing W, Four D, Generation TH, For Z, Effect TOK, Absorption T. Experimental Techniques and Details : Degenerate Four Wave Mixing and Z-scan. *Time*.

46	Koletar MM, Dorr A, Brown ME, McLaurin JA, Stefanovic B. Refinement of a chronic cranial window implant in the rat for longitudinal in vivo two–photon fluorescence microscopy of neurovascular function. *Sci Rep* 2019; **9**. doi:10.1038/s41598-019-41966-9.

47	Mostany R, Portera-Cailliau C. A craniotomy surgery procedure for chronic brain imaging. *Journal of Visualized Experiments* 2008. doi:10.3791/680.

48	Neethling PH. Determining non-linear optical properties using the Z-scan technique. 2005.http://scholar.sun.ac.za/handle/10019.1/2094.

49	Bojtár M, Kormos A, Kis-Petik K, Kellermayer M, Kele P. Green-Light Activatable, Water-Soluble Red-Shifted Coumarin Photocages. *Org Lett* 2019; **21**. doi:10.1021/acs.orglett.9b03624.

50	Rumi M, Ehrlich JE, Heikal AA, Perry JW, Barlow S, Hu Z *et al.* Structure - Property relationships for two-photon absorbing chromophores: Bis-donor diphenylpolyene and bis(styryl)benzene derivatives. *J Am Chem Soc* 2000; **122**. doi:10.1021/ja994497s.

51	Wawrzynczyk D, Szeremeta J, Samoc M, Nyk M. Optical nonlinearities of colloidal InP@ZnS core-shell quantum dots probed by Z-scan and two-photon excited emission. *APL Mater* 2015; **3**. doi:10.1063/1.4935748.

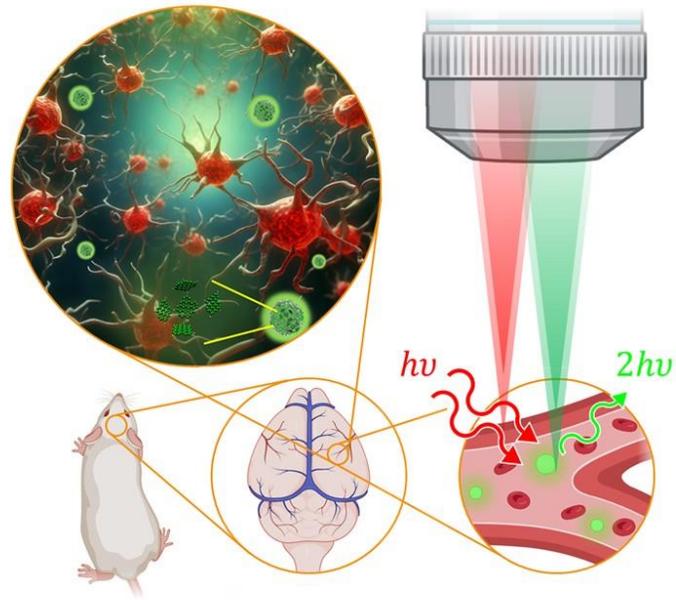

TOC
Summarized description of the paper content, visualization of vaterite particles loaded with vaterite in two-photon fluorescence technique.